\documentclass[aps,prd,showpacs,preprintnumbers,twocolumn,groupedaddress,nofootinbib]{revtex4}
\usepackage{graphicx}
\usepackage{color}
\usepackage{latexsym}
\usepackage{booktabs}
\usepackage{multirow}
\usepackage{amsmath, footnote, epsfig}
\pdfoutput=1

\def\beq{\begin{equation}}
\def\eeq{\end{equation}}
\def\bey{\begin{eqnarray}}
\def\eey{\end{eqnarray}}

\def\lsim{\mathrel{\raise.3ex\hbox{$<$\kern-.75em\lower1ex\hbox{$\sim$}}}}
\def\gsim{\mathrel{\raise.3ex\hbox{$>$\kern-.75em\lower1ex\hbox{$\sim$}}}}

\begin{document}

\title{Known Radio Pulsars Do Not Contribute to the Galactic Center Gamma-Ray Excess}
\author{Tim Linden}
\affiliation{Kavli Institute for Cosmological Physics, University of Chicago, Chicago, IL, 60637}
\affiliation{Center for Cosmology and AstroParticle Physics (CCAPP) and Department of Physics, The Ohio State University Columbus, OH, 43210 }
\begin{abstract}
Observations using the Fermi Large Area Telescope (Fermi-LAT) have found a significant $\gamma$-ray excess surrounding the center of the Milky Way (GC). One possible interpretation of this excess invokes $\gamma$-ray emission from an undiscovered population of either young or recycled pulsars densely clustered throughout the inner kiloparsec of the Milky Way. While these systems, by construction, have individual fluxes that lie below the point source sensitivity of the Fermi-LAT, they may already be observed in multiwavelength observations. Notably the Australia Telescope National Facility (ATNF) catalog of radio pulsars includes 270 sources observed in the inner 10$^\circ$ around the GC. We calculate the $\gamma$-ray emission observed from these 270 sources and obtain three key results: (1) point source searches in the GC region produce a plethora of highly significant $\gamma$-ray ``hotspots", compared to searches far from the Galactic plane, (2) there is no statistical correlation between the positions of these $\gamma$-ray hotspots and the locations of ATNF pulsars, and (3) the spectrum of the most statistically significant $\gamma$-ray hotspots is substantially softer than the spectrum of the GC $\gamma$-ray excess. These results place strong constraints on models where young pulsars produce the majority of the $\gamma$-ray excess, and disfavors some models where milli-second pulsars produce the $\gamma$-ray excess.
\end{abstract}
\maketitle

\section{Introduction}
\label{sec:introduction}

The center of the Milky Way Galaxy (GC) is expected to provide the brightest signal from dark matter annihilation in the universe, making it a key target for $\gamma$-ray searches probing the nature of the dark matter particle. However, the GC also contains myriad astrophysical backgrounds, including point-like emission from populations of supernova remnants, both young and mature pulsars, as well as diffuse emission from astrophysical pion-decay, bremsstrahlung, and inverse-Compton scattering. Efforts to utilize GC observations in order to probe fundamental physics must carefully disentangle these bright astrophysical backgrounds.

Intriguingly, observations by Goodenough and Hooper~\citep{Goodenough:2009gk, Hooper:2010mq}, noted the existence of an excess in GeV $\gamma$-ray emission centered at the GC. Subsequent studies further detailed the key features of this excess, characterizing it as a ``bump" in $\gamma$-ray flux at an energy between 1--3~GeV, with a low-energy spectrum that is harder than expected from astrophysical pions~\citep{Hooper:2011ti, Abazajian:2012pn, Gordon:2013vta, Macias:2013vya, Abazajian:2014fta}. Moreover, the excess morphology was shown to be consistent with expectations from annihilating dark matter. Work by Hooper \& Slatyer~\citep{Hooper:2013rwa} expanded the region of interest (ROI) of these observations, utilizing a novel background subtraction technique to produce evidence of extended $\gamma$-ray emission out to $\sim$15$^\circ$ from the GC. Daylan et al.~\citep{Daylan:2014rsa} first integrated these two experimental ROIs, using the combined power of each approach to demonstrate that the excess: (1) is approximately spherically symmetric around the dynamical center of the Galaxy, (2) has an emission profile consistent with a power-law between approximately 15 -- 1500 pc from the GC, and (3) is spectrally consistent in ROIs that are of varying distance from the GC. Most recently, Calore et al.~\citep{Calore:2014xka} showed that these key parameters of the $\gamma$-ray excess remain resilient to changes in the astrophysical diffuse emission models.

Several classes of models have been put forward to explain the observed features of the $\gamma$-ray excess. Models attributing the GeV excess to dark matter annihilation have been particularly popular, due to the simplicity of fitting the intensity, spectrum and morphology of the $\gamma$-ray excess with the theoretically predicted cross-section, annihilation spectrum and density profile of weakly-interacting dark matter~\citep[see e.g.][]{Berlin:2014tja, Agrawal:2014una, Alves:2014yha, Abdullah:2014lla, Izaguirre:2014vva}. Models attributing the excess to either hadronic~\citep{Carlson:2014cwa} or leptonic~\citep{Petrovic:2014uda,Cholis:2015dea} outbursts from the central black hole have been motivated primarily by the known variability of the GC region, and the existence of non-steady state emission sources such as the Fermi bubbles. Finally, models of pulsar emission have been posited due to the spectral similarilty of the GeV excess to the population of observed $\gamma$-ray pulsars~\citep{Abazajian:2010zy, Abazajian:2012pn, Yuan:2014yda, Petrovic:2014xra, O'Leary:2015gfa}. While the majority of recent works attribute pulsar emission in the GC region to a population of mature MSPs, recent analyses have also explained the GC excess with a population of young pulsars in the GC region~\citep{O'Leary:2015gfa}.

However, it is difficult to reconcile the large number of pulsars necessary to explain the GC excess with the lack of observed $\gamma$-ray pulsars near the GC~\citep{Hooper:2013nhl, Cholis:2014noa, Cholis:2014lta}. Specifically, Cholis et al. \citep{Cholis:2014lta} argue that pulsar models for the GC excess would be expected to contain 226.9$^{+91.2}_{-67.4}$ pulsars with a $\gamma$-ray luminosity above 10$^{34}$~erg~s$^{-1}$ and 61.9$^{+60.2}_{-33.7}$ with a $\gamma$-ray luminosity above 10$^{35}$~erg~s$^{-1}$ within the inner 10$^\circ$ around the GC, though we note that these estimates could be decreased if there are systematic overestimates in the distances to known MSPs~\citep{Brandt:2015ula}. While these sources may potentially remain hidden if they reside in the inner few degrees around the GC, where the astrophysical diffuse emission is particularly bright, a large population is expected to be observed in regions farther from the GC. However, this analysis assumes that the population of pulsars in the GC region have similar characteristics as those found in the galactic plane. Observational studies are thus necessary in order to conclusively find or rule out pulsar contributions to the GeV excess.

Very recently, two separate groups have provided observational evidence for significant point-source emission in the GC region. One analysis by Bartels et al.~\citep{Bartels:2015aea} examines sources observed in the Fermi-LAT Collaborations Third Catalog of $\gamma$-ray point sources (3FGL)~\citep{TheFermi-LAT:2015hja}. They argue that a number of currently unassociated point sources near the GC have spectra which are consistent with pulsar emission. Employing a wavelet analysis in order to analyze the characteristics of sources not bright enough to reside in the 3FGL catalog, they additionally find evidnence that numerous sub-threshold point sources exist. Combining these two observations, they argue that the currently observed 3FGL sources constitute only the tip of the iceburg, and a significant population of sub-threshold pulsars exists throughout the inner galaxy. A second, detailed analysis by \citep{Lee:2015fea} examines the photon flux in the inner 30$^\circ$ around the GC and utilizes a one-point non-Poissonian template fit to study the emission properties of the underlying $\gamma$-ray emitters. They find that the data contain more high-flux pixels than expected from a smooth emission source, such as dark matter annihilation. Instead, the template fit appears consistent with an excess produced by a population of sub-threshold point sources. Intriguingly, the majority of these sources appear to exist at fluxes just below the Fermi-LAT point source sensitivity, indicating that future observations may successfully detect these objects as individual $\gamma$-ray point sources.

While these new $\gamma$-ray analyses offer an intruiging new methodology for probing the morphology of the GC $\gamma$-ray excess, they are, at present, not able to convincingly distinguish between true point source emission and residuals stemming from the mis-subtraction of diffuse structures in the inner galaxy. This work presents a possible method to distinguish these two possibilities. Specifically, the existence of a population of statistically significant (yet still sub-threshold) hotspots allows for a statistical comparison of these hotspots with multiwavelength source catalogs in order to determine the underlying emission source. In the case of pulsars, there are numerous radio catalogs of the GC region. Here, we employ the Australia Telescope National Facility (ATNF) catalog\footnote{http://www.atnf.csiro.au/people/pulsar/psrcat/}~\citep{Manchester:2004bp}, which includes 270 known radio pulsars within 10$^\circ$ of the GC. If these pulsars were to statistically coincide with $\gamma$-ray hotspots observed by the Fermi-LAT, it would serve as smoking gun evidence indicating a pulsar origin for the GC excess.


In this paper, we search for sub-threshold $\gamma$-ray point sources at the positions of known ATNF pulsars near the GC, comparing the distribution of $\gamma$-ray fluxes found at these sky positions with that of equivalent null sky positions. Our analysis finds no statistical correlation between ATNF pulsars and $\gamma$-ray emission, ruling out scenarios where known radio pulsars contribute significantly to the GC excess. We additionally calculate the average spectrum of the brightest point source hotspots in the GC, finding that they contain significantly more low-energy emission than is observed in models of the GC excess, which disfavors some models where point sources dominate the GC excess.

\section{Models}\label{sec:model}

\subsection{ATNF Pulsar Catalog}
In this paper, we examine the population of radio pulsars which are not currently associated with any 3FGL point source. We utilize the ATNF catalog, which currently includes a population of 2405 observed radio pulsars with a morphological distribution that is strongly peaked near the galactic plane for both physical and observational reasons. Of these pulsars, 270 are located within 10$^\circ$ of the GC. In this analysis we combine any pulsars found within the same globular cluster, since they have angular separations much smaller than the point-spread function of the Fermi-LAT. This leaves us 201 independent sky locations containing an ATNF source. We then remove from our analysis 16 ATNF pulsars that lie within 0.1$^\circ$ of any $\gamma$-ray source listed within the Fermi-LAT collaboration 3FGL catalog~\citep{TheFermi-LAT:2015hja}. Of these 16 pulsars, 15 lie within 0.1$^\circ$ of a single 3FGL point source, and the ATNF source J1745-2900 lies within 0.1$^\circ$ of two 3FGL sources. Of the 17 Fermi-LAT sources located near an ATNF pulsar, 13 are already associated with either individual pulsars, pulsar wind nebulae, or globular clusters (3FGL J1701.2-3006, 3FGL J1732.5-3130,  3FGL J1741.9-2054, 3FGL J1746.3-2851c, 3FGL J1746.8-3240, 3FGL J1747.2-2958, 3FGL J1748.9-2021, 3FGL J1748.0-2447, 3FGL J1750.2-3704, 3FGL J1803.1-2147, 3FGL J1809.8-2332, 3FGL J1823.7-3019, 3FGL J1824.6-2451), two are currently unassociated (3FGL J1736.5-2839 and 3FGL J1745.3-2903c), one is associated with a nearby supernova (3FGL J1741.1-3053), and one with either a supernova or PWN (3FGL J1745.6-2859c). We note that the majority of these sources are already modeled (or masked) in analyses of the GC $\gamma$-ray excess, and thus do not affect the results of previous works. 

This leaves us 185 ATNF pulsars that are at least 0.1$^\circ$ removed from any known 3FGL source, decreasing the impact of 3FGL source mis-modeling on the $\gamma$-ray intensity at each ATNF pulsar location.  We note that the choice of a 0.1$^\circ$ cut to ensure the separation of $\gamma$-ray point sources may appear ambitious, and we will comment on this choice in Section~\ref{sec:fermimodels}. Here, we note only that placing larger cuts on the separation between ATNF pulsars and Fermi point sources would create systematic biases in the morphology of remaining ATNF pulsars, preferentially moving sources near the GC and Galactic plane. 

Many of these 185 pulsars lie in the direction of the GC, but have distance measurements that make them unlikely to be located within the GC region. If these systems were to be $\gamma$-ray bright, they may still contribute to the GC excess, due to the lack of distance information in the $\gamma$-ray signal. On the other hand, the $<$0.05$^\circ$ offset between the peak of the $\gamma$-ray excess and the position of Sgr A* strongly indicates that the majority of the GC excess is, in fact, dynamically centered at the GC~\citep{Daylan:2014rsa}. Of the 185 ATNF pulsars in our study, only 41 have best fit distances falling between 6.0 -- 11~kpc from the sun, which would place them in the GC region. However, these distance measurements are based on free-electron models of the galaxy~\citep{Taylor:1993my}, and have considerable uncertainties, especially for sources close to the GC~\citep{2002astro.ph..7156C}. It is thus possible that the subset of pulsars corresponding to the dynamical GC is either significantly larger or smaller than assumed here. In what follows, we arbitrarily denote ``GC" ATNF pulsars to be those with a distance measurement falling between 6.0 -- 11~kpc from the sun, noting that this selects a statistical sub-sample of pulsars which are more likely to be affiliated with the GC than our full sample.

Finally, we note that the majority of ATNF pulsars are young systems. Of the 185 ATNF pulsars in our analysis, 17 exhibit periods smaller than 30~ms, and are designated as MSPs. Of these 17 MSP systems, 3 have distance measurements which place them near the dynamical center of the galaxy. Thus, ATNF pulsar searches are significantly more sensitive to scenarios where young pulsars contribute significantly to the GeV excess, such as those put forward by~\citep{O'Leary:2015gfa}. However, it is worth noting that the lack of MSPs observed in ATNF observations of the GC region may significantly constrain models attributing the GC excess to populations of mature pulsars.

\subsection{Fermi Data Analysis}
\label{sec:fermimodels}

In order to calculate the $\gamma$-ray emission coincident with ATNF pulsars in the GC region, we utilize 6.5 years of Fermi-LAT data\footnote{MET Range: 239557417-452717733}, including P7V6\_REP events which are converted both in the front and back of the Fermi-LAT instrument. We process front and back converting events separately throughout our analysis chain, but combine the resulting likelihood profiles in order to calculate the significance of each source. However, when considering source spectra, we consider only front-converting events due to their superior angular resolution, which decreases the effect of diffuse background mis-modeling on the point source properties. We place several important cuts on our event selection, restricting ourselves to photons belonging to the SOURCE class which arrive at a zenith angle smaller than 100$^\circ$ and that are observed when the instrument is oriented with a rocking angle smaller than 52$^\circ$ compared to the Earth's zenith. We additionally remove events taken when the instrument is not in both science mode and survey mode, and events recorded while the instrument is passing through the South Atlantic Anomaly. 

In order to determine the significance and spectrum of any point source coincident with each ATNF pulsar, we utilize a likelihood algorithm similar to that employed in dark matter searches for dwarf spheroidal galaxies~~\citep{Ackermann:2013yva, Hooper:2015ula}. Specifically, we bin photons into a 10$^\circ \times 10^\circ$ box centered on the position of the posited point source, dividing the box into 200~$\times$~200 spatial bins and 18 energy bins logarithmically spaced between 100~MeV and 100~GeV. We first fit the data over all 18 energy bins, allowing the normalization and spectrum of the ATNF pulsar, all nearby point sources, and all diffuse sources to float\footnote{Throughout this work we adopt the Pass 7 Reprocessed Diffuse Model (gll\_iem\_v05\_rev1.fit). designed by the Fermi-LAT collaboration for point source searches using Reprocessed Pass 7 Data.} We then fix the source spectrum and normalization of all sources besides the pulsar under investigation, and then calculate the likelihood profile obtained by varying the normalization of the pulsar in each energy bin independently. In analyzing the test statistic of the source, we must assume a spectral model, and we take each putative pulsar to be represented by a LogParabola spectrum \footnote{The LogParabola spectrum employed in this work is given by: $\frac{dN}{dE}~=~N_0~\left(~\frac{E}{E_b}\right)^{-\alpha~+ \beta~log(E/E_b)}$}, with N$_0$ allowed to float freely, -3~$\leq$~$\alpha$~$<$~0  and -0.5~$\leq$~log$_{10}(\beta)$~$\leq$~1.0.  While we note that this allows for very soft spectral features to fit the $\gamma$-ray data, we have checked that increasing the minimum values of $\alpha$ and $\beta$ have little effect on the results of our analysis.

\begin{figure}
\includegraphics[width=250pt]{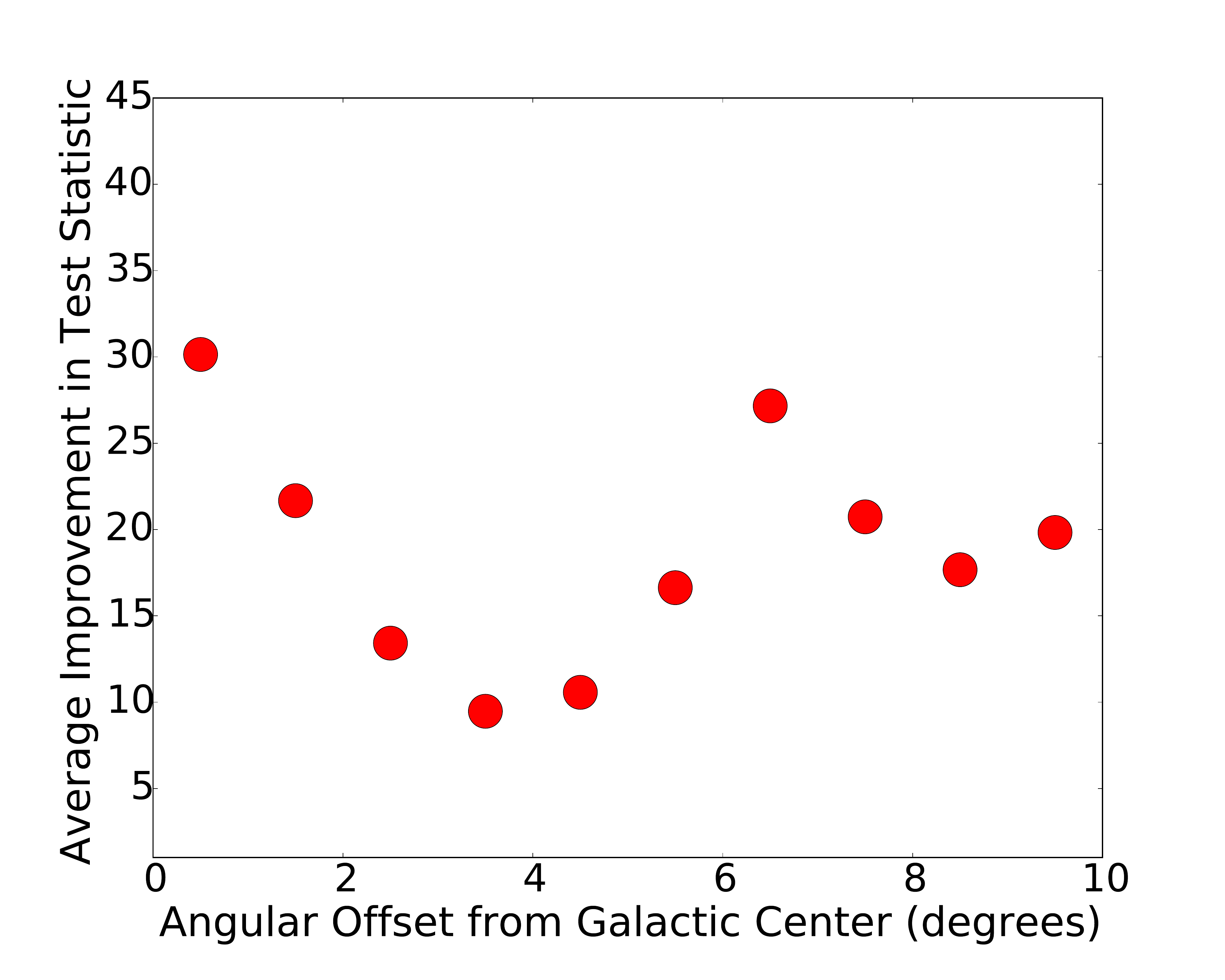}
\caption{\label{fig:fluctuations_vs_radius} The average improvement in Test Statistic (TS) obtained when point sources are added at various sky positions centered on either the position of an ATNF pulsar, or on a ``mirrored" sky position at an identical radial distance from the GC. The resulting TS distribution is shown as a function of the distance between the random sky position and the dynamical center of the galaxy. Error bars are not provided due to the statistical interdependence of nearby sky positions.}
\end{figure}

The typical threshold for the ``detection" of a Fermi-LAT point-source has typically been assumed to be TS~=~25, which statistically translates to a 5$\sigma$ detection in cases where the background is described to the level of Poisson noise. In the more realistic case where systematic uncertainties contribute to $\gamma$-ray residuals in addition to statistical uncertainties, a TS~=~25 source has been determined to more closely correlate with a $\approx$4$\sigma$ detection. However, these rule-of-thumb values are unlikely to be accurate in the dense GC region. Specifically, the bright $\gamma$-ray flux from diffuse astrophysical emission can mimic point source emission, and produce a significant population of fake $\gamma$-ray point source emitters. It is difficult to utilize mock data in order to estimate the magnitude of this effect, since diffuse models lack the resolution to represent the many features of the GC region. 

\begin{figure*}
\includegraphics[width=500pt]{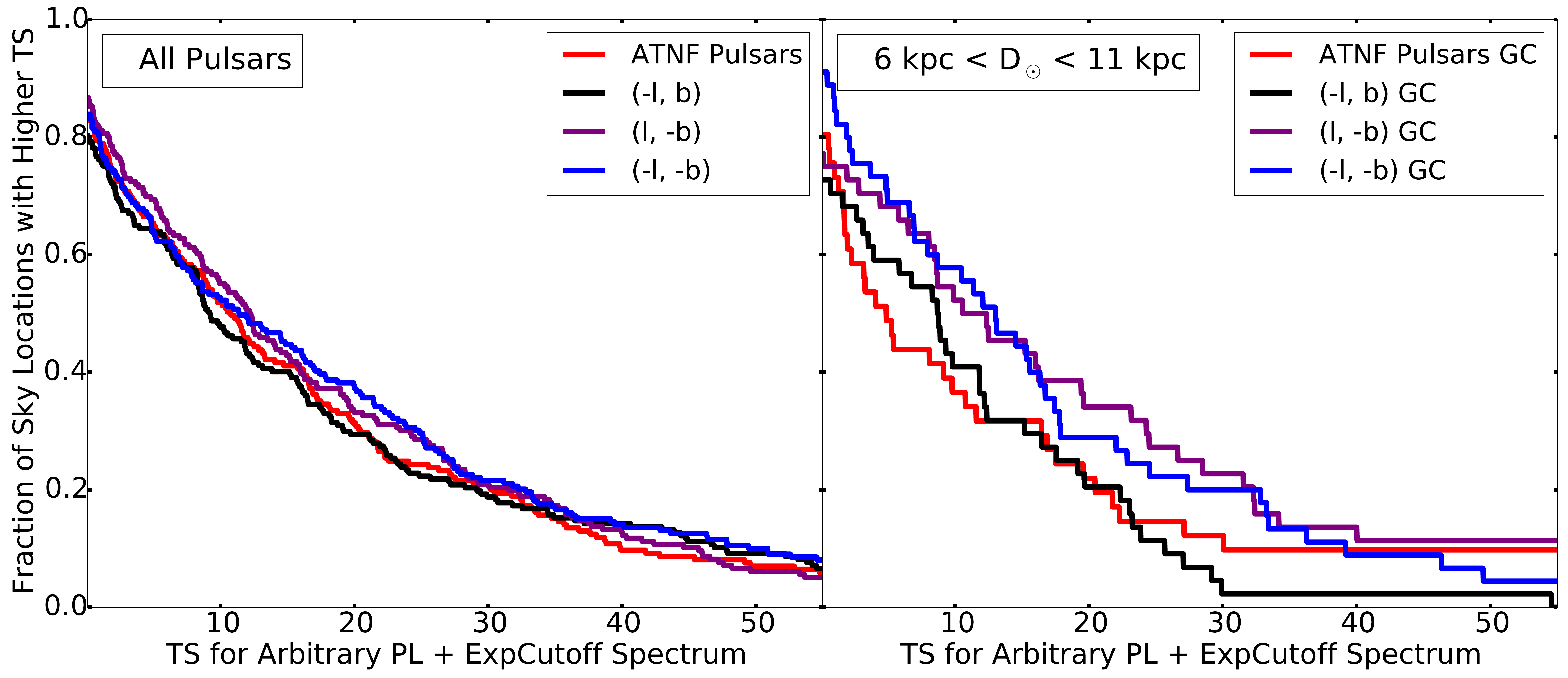}
\caption{\label{fig:TSdistribution} The distribution of TS values obtained when point-sources are added at sky positions corresponding to the 185 ATNF pulsars located within 10$^\circ$ of the GC (left) and the sub-sample of 41 ATNF pulsars which also have best fit distance measurements between 6.0-11.0~kpc from the solar position (right). The $\gamma$-ray source significance is evaluated utilizing the Fermi-LAT tools as described in Section~\ref{sec:model}. The results are compared with the distribution of TS values obtained when Fermi point sources are added to ``mirrored" blank sky locations with similar background characteristics to the locations of ATNF pulsars.}
\end{figure*}

We caution the reader that there are several important distinctions between the analysis method employed here, and that used in the 3FGL catalog employed by the Fermi-LAT collaboration~\citep{TheFermi-LAT:2015hja}, which make it difficult to make straightforward comparisons between the results shown here and catalog sources. For example, (1) this analysis utilizes 6.5 years of data rather than 4 years of data, (2) this analysis does not allow for energy-dependent alterations in the diffuse background normalization, nor does it make ``\emph{ad hoc}" adjustments to the log-likelihood contribution based on the diffuse model uncertainties in the region. We do not argue that the techniques applied here are in any way superior to those employed by the 3FGL catalog, but note that they make direct comparisons difficult. Most importantly, a TS~=~25 excess calculated here does not correspond to a 4$\sigma$ point source detection.

In order to calculate the significance of ATNF pulsar contributions to the GC excess, we instead compare the ATNF sample with a data driven null sample by taking advantage of the axisymmetric nature of the GC diffuse emission. In addition to analyzing the $\gamma$-ray point source emission at the position ($\ell$, $b$) of each known ATNF pulsar, we also place fake pulsar sources at the positions (-$\ell$, $b$), ($\ell$, -$b$), and (-$\ell$, -$b$), again removing any fake pulsar sources which are within 0.1$^\circ$ of a 3FGL source. While these fake pulsar positions reside in regions with different diffuse backgrounds for each individual pulsar, the statistical sample of these sky positions is expected to have the same characteristics as the actual ATNF population. By comparing the population of ATNF pulsars with these null positions, we can look for statistical evidence of a $\gamma$-ray excess stemming from the ATNF pulsar population.

We note that some ATNF pulsar locations may lie in close proximity to known $\gamma$-ray source. Specifically, our analysis removes only ATNF pulsars which lie within 0.1$^\circ$ of a 3FGL source. This is smaller than the positional accuracy of many 3FGL source locations\footnote{ATNF pulsar locations have significantly better positional reconstructions, and positional errors in radio source locations are irrelevant to our analysis.}. This could potentially produce two systematic biases. First, ATNF pulsars may acquire significant photon fluxes from a nearby 3FGL source with which it is not affiliated. However, we note that the 3FGL source is not removed from our analysis. Instead, they are allowed to float in normalization and spectrum, and are able to soak up $\gamma$-ray excesses at the 3FGL source location. Moreover, the size of the positional error cone is irrelevant, because the 3FGL source position is directly calculated from a largely overlapping set of Fermi-LAT $\gamma$-ray data. In this case, we expect that the 3FGL source will properly acquire the majority of the point source flux, and insignificant leakage to the nearby ATNF source will occur. 

A second scenario involves an ATNF source which is, in fact, affiliated with a nearby 3FGL source, but lies farther than 0.1$^\circ$ from the best-fit 3FGL source location. In this case, there will be a suppression of the $\gamma$-ray flux of the true ATNF source due to the existence of the spurious 3FGL source. We note two facts that mitigate the effect of this systematic on our analysis. First, none of the 3FGL sources in our ROI that have been previously associated with an ATNF pulsar have an angular offset greater than 0.03$^\circ$, which is significantly smaller than our ROI. Second, we find 20 ATNF pulsars which lie between 0.1$^\circ$ and 0.3$^\circ$ of a 3FGL source, a result which is comparable to the 18.3 chance associations observed for each of our ``mirrored" ATNF blank sky locations. This implies that very few real associations exist, that have not been taken into account with our 0.1$^\circ$ sky cut.

\section{The Flux Distribution of ATNF Pulsars}
\label{sec:results}

In order to determine the significance of ATNF sources observed in the GC region, we first calculate the distribution of TS values obtained in blank sky locations throughout our ROI. In Figure~\ref{fig:fluctuations_vs_radius} we calculate the average TS value of our pulsar and blank sky locations as a function of their angular separation from the GC, and note two key results. First, the average TS value in this region significantly exceeds expectations from either Poisson fluctuations ($\overline{TS}$~=~0.5) or from null sky observations in the high latitude sky ($\overline{TS}$~=~0.98~\citep{Hooper:2015ula}). This result is compatible with the work of \citep{Bartels:2015aea, Lee:2015fea}, showing that point sources placed in the GC can pick up significant $\gamma$-ray fluxes. However, we caution that this does not conclusively demonstrate that bright point sources reside in the GC region. Since our analysis only tests the improvement in a model after a point source degree of freedom is added, it is unable to distinguish between true point sources and errors in the modeling of the diffuse background. Second, we note that the average point source TS varies significantly within the ROI of our study, making it difficult to calculate the significance of any excess through a comparison to the average distribution of TS values within the ROI. 

\begin{figure*}
\includegraphics[width=500pt]{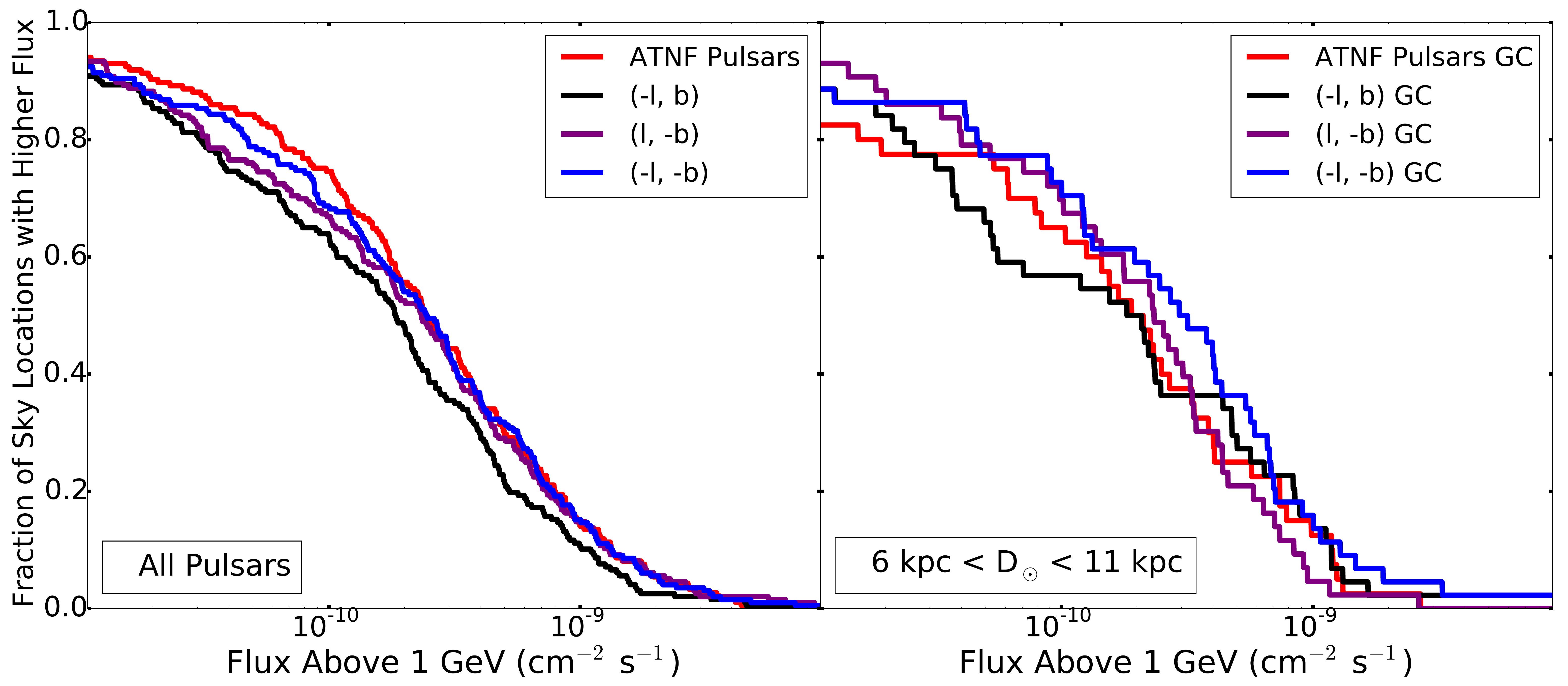}
\caption{\label{fig:fluxdistribution} Same as Figure~\ref{fig:TSdistribution}, but with the distributions shown in terms of the $\gamma$-ray flux above 1 GeV, rather than the Test Statistic of the $\gamma$-ray source. For the calculation of source fluxes, we restrict our analysis to consider only events which convert in the front of the Fermi-LAT telescope.}
\end{figure*}

The strong dependence of the point source TS on the angular separation between a source and the galactic center again motivates the ``axisymmetric blank sky location" test described in Section~\ref{sec:model}. In Figure~\ref{fig:TSdistribution} (left) we show the TS distribution for sources placed at the position of ATNF pulsars, as well as at the `mirrored" sky positions. This reveals two primary results of our analysis. First, we again note the overabundance of high TS sources. Nearly 40\% of ATNF pulsar locations produce a TS exceeding 10, while this is true for only $\sim$1\% of blank sky locations in the high-latitude sky~\citep{Hooper:2015ula}. We note that nearly 20\% of these sources have TS values exceeding 25, which might appear at odds with their absence from the 3FGL catalog. However, there are a multitude of technical differences between this analysis and that employed by the 3FGL catalog which make a straightforward comparison impossible, as described in Section~\ref{sec:model}.

More importantly, we note that the TS distributions of ``mirrored" sky locations are statistically equivilent to the TS distributions of real ATNF pulsars, implying that even high-TS ATNF pulsars are unlikely to represent physical $\gamma$-ray emitters. In Figure~\ref{fig:TSdistribution} (right), we find that this result is unchanged when we restrict our analysis to the 41 ATNF pulsars with distance measurements placing them near the dynamical center of the galaxy. Quantitatively, we find that 31.35\% (58/185) of ATNF locations contain a hotspot with TS$>$20, compared to 33.44\% (198/592) of mirrored sky locations. Since mirrored sky locations contain no ATNF pulsar contribution, this constrains the number of ATNF pulsars contributing to the TS$>$20 hotspots to be smaller than 10.36 at 95\% confidence. Constraining ourselves to pulsars in the GC region, we similarly find that 21.95\% (9/41) ATNF pulsar locations produce a TS$>$20 hotspot compared to 27.82\% (37/133) of null sky locations. This constrains the number of GC ATNF pulsars contributing to the TS$>$20 hotspots to be smaller than 3.65 at 95\% confidence. 

In Figure~\ref{fig:fluxdistribution} we show the same populations of ATNF pulsars and mirrored sky positions plotted as a function of the $\gamma$-ray flux, and examining only the $\gamma$-ray flux above 1 GeV (similar to both the 3FGL catalog and the analysis of \citep{Lee:2015fea}). We note that our typical sky position contributes a $\gamma$-ray flux of a few~$\times$~10$^{-10}$~cm$^{-2}$s$^{-1}$, which is similar to both the sensitivity level of the 3FGL catalog, and to the $\gamma$-ray fluxes of point source hotspots observed by \citep{Lee:2015fea}. Since point source explanations for the $\gamma$-ray excess require a population of several hundred sources with fluxes just below the Fermi-LAT point source detection threshold~\citep{Lee:2015fea}. The lack of any observed excess coincident with ATNF pulsar locations thus makes it unlikely that these sources contribute to the GC $\gamma$-ray excess or the point source hotspots identified by \citep{Lee:2015fea}.

More quantitatively, we can place constraints on the total emission from ATNF pulsars by examining the flux distributions shown in Figure~\ref{fig:fluxdistribution}. Utilizing the average pulsar spectrum obtained by~\citep{Cholis:2014noa}, we note that a pulsar with a $\gamma$-ray flux of 10$^{34}$ erg~s$^{-1}$ above 100 MeV and located at a distance 8~kpc from the solar position contributes a $\gamma$-ray flux above 1 GeV of 1.9~$\times$10$^{-10}$ ph~cm$^{-2}$s$^{-1}$. While the analysis of \citep{Cholis:2014lta} requires a population of 226.9$^{+91.2}_{-67.4}$ such systems in order to explain the GC excess, our analysis constrains the total number of ATNF systems above this flux-threshold to be below 14.4 at 95\% confidence. Considering the sub-population of pulsars with distance measurements placing them in the GC region, we constrain the number of ATNF systems above this flux threshold to be less than 4.76 at 95\% confidence. We thus constrain ATNF pulsars from producing more than $\sim$10\% of the GC excess emission. We stress that in all cases, our results are consistent with the null hypothesis, that ATNF pulsars contribute no $\gamma$-ray emission to the $\gamma$-ray excess.

\section{The Spectrum of Point Sources near the GC}
\label{sec:spectrum}

\begin{figure*}
\includegraphics[width=500pt]{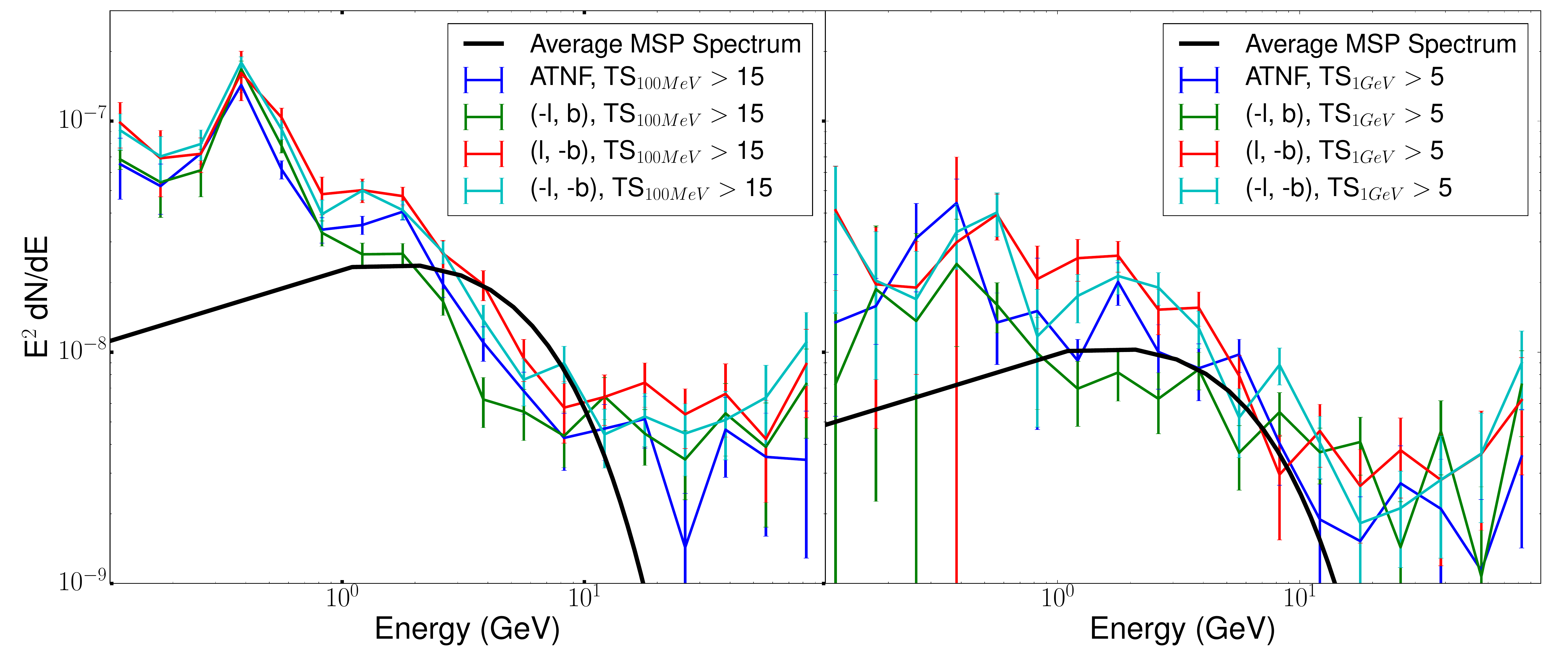}
\caption{\label{fig:ATNF_spectrum} The average $\gamma$-ray spectrum of ATNF pulsars observed with TS~$>$~15 over the full energy range (left), and TS~$>$~5 for photons over 1~GeV (right), compared to the average spectrum of ``mirrored" sky locations passing the same TS cut. We compare the resulting spectrum with the Average $\gamma$-ray spectrum of observed MSPs as determined by~\citep{Cholis:2014noa}. The spectral analysis is carried out utilizing only front-converting events, in order to remove systematic errors stemming from the poor angular resolution of back-converting events at low-energies.} 
\end{figure*}

While the above results indicate that ATNF pulsars do not provide a substantial contribution to the intensity of the $\gamma$-ray excess, it is worth checking whether some subset of high TS $\gamma$-ray point sources have spectra similar to that of $\gamma$-ray pulsars (and by extension to the GC excess). This might indicate that point sources near the GC contribute heavily to the GC excess spectrum, even if the point sources are not strongly correlated to ATNF pulsar locations.  In all spectral calculations that follow, we restrict our analysis to front-converting events, which have a superior angular resolution in the $\sim$~1~GeV energy range where the characteristic pulsar spectrum peaks. In Figure~\ref{fig:ATNF_spectrum} (left) we select the most significant ATNF source positions by placing an arbitrary cut including only pulsars with TS$>$15. We also show the average $\gamma$-ray spectrum of mirrored sky locations passing the same TS cut. We compare these resulting $\gamma$-ray spectra with the average spectra of MSPs calculated by~\citep{Cholis:2014noa}.

We note two results which disfavor scenarios where a substantial fraction of even the brightest point sources observed in our analysis correspond to the GC excess. First, we find that the $\gamma$-ray spectrum from ``mirrored" pulsars is statistically idistinguishable to that from true ATNF pulsar locations, disfavoring any ATNF contribution to the GC excess. Second, we find that the average spectrum of these point sources is far softer than observed from $\gamma$-ray pulsars~\citep{Cholis:2014noa}, with a smoothly falling spectrum of approximately E$^{-2.5}$. This disfavors scenarios where a significant fraction of any high-TS sky locations in this analysis contribute meaningfully to the $\gamma$-ray excess.

In the analysis of Lee et al.~\citep{Lee:2015fea}, the significance of $\gamma$-ray hotspots is calculated from photons in the energy range 1.893~--~11.943~GeV. This restriction is necessary to diminish the sensitivity of the non-Poissonian template fit to the rapidly changing instrumental point-spread function at low energies. In order to create results more comparable with this work, in Figure~\ref{fig:ATNF_spectrum} (right), we repeat the above analysis, considering only pulsars with TS$>$5 at an energy above 1~GeV. Even after imposing a strong selection effect for GeV photons (by considering only the systems which are brightest in this energy range), the average spectrum of these sources remains softer than expected for physical pulsars.

The results of Figure~\ref{fig:ATNF_spectrum} additionally cast doubt on the ability of any point source population to fit the spectrum and intensity of the GC excess. Pulsar models of the GC excess require several hundred bright pulsars to produce emission in the GC region. While there do appear to be locations in the GC where point sources can produce considerable emission, they primarily adopt power-law spectrum which do not reflect the global properties of the GC excess. However, it is currently unclear whether the stacked spectral fits of individual $\gamma$-ray point sources produces an accurate representation for the total spectrum from an ensemble of point source emitters. One could imagine that the spectra of individual point sources in this region of the sky have spectra which are highly degenerate in a way which systematically effects their stacked spectra. Further investigation of the spectral properties of point sources in the GC and their effect on the GC excess will be left to an upcoming publication~\citep{nextpaper}.

\section{Placing Limits on the Pulsar Contribution}
\label{sec:limits}

Our analysis shows that there is no correlation between the position of $\gamma$-ray hotspots and the observed locations of ATNF pulsars. This is unfortunate, as a clear correlation between ATNF pulsars and $\gamma$-ray hotspots would be a bulletproof indication of pulsar contributions to the GC excess. It is worth considering whether this null result places any constraints on pulsar contributions to the GC excess, or whether ATNF pulsars constitute too small a percentage of the total GC pulsar population in order to expect any overlap between the positions of $\gamma$-ray and ATNF pulsars. 

In order to determine the expected overlap between $\gamma$-ray pulsars in the GC region and ATNF pulsars, we compare the population of 3FGL sources in the galactic plane with the ATNF population. We define two analysis regions: the GC region is defined as $|\ell|<$10$^\circ$, $|b|<$10$^\circ$, and the positive and negative sidebands regions are defined using 10$^\circ<|\ell|<$30$^\circ$, $|b|<$10$^\circ$, with $\ell$ positive and negative respectively. Removing 3FGL sources that are associated with non-pulsar objects (but keeping those associated with globular clusters and PWN), we find 74 3FGL sources in the GC region, compared to 60 in the negative sideband and 40 in the positive sideband. We consider a source (associated or not) to be correlated with an ATNF pulsar if it resides within 0.2$^\circ$ of an ATNF pulsar source\footnote{This corresponds approximately to the 95\% containment region of 3FGL point sources, and leaves a chance coincidence of $\sim$8\% for any individual 3FGL source}. We find that 23 (31\%) of GC 3FGL sources are correlated with an ATNF source, compared to 13(22\%) of negative sideband sources and 16(40\%) of positive sideband sources. This indicates that pulsars in the GC region are similarly correlated with ATNF pulsars as those in the surrounding galactic plane - and offers evidence that the source classes producing $\gamma$-ray emission in the GC region are similar to those in the rest of the galactic plane.

If the probability that a $\gamma$-ray pulsar has an ATNF counterpart is independent of the $\gamma$-ray flux, this observation would provide strong evidence not only against ATNF pulsar interpretations of the $\gamma$-ray excess, but against any pulsar explanation for the $\gamma$-ray excess.  Removing the 8\% chance coincidence between ATNF pulsars and 3FGL sources, we would naively expect approximately one-quarter of $\gamma$-ray pulsars to correspond to sources in the ATNF catalog. Noting that approximately 200 systems with a $\gamma$-ray luminosity exceeding 10$^{34}$~erg~s$^{-1}$ are necessary to explain the $\gamma$-ray excess~\citep{Cholis:2014lta}, this would predict approximately 50 systems correlated with ATNF pulsars. This scenario is clearly ruled out by the analysis of Section~\ref{sec:results}. However, it is difficult to determine whether a correlation between the $\gamma$-ray luminosity and radio detectability of pulsars exists, due to the multiple selection effects in both blind and targeted pulsar searches. 

Interestingly, recent analyses have argued that the radio loudness of pulsars is uncorrelated with their $\gamma$-ray emission properties~\citep{Marelli:2015vsa}. In the case of MSPs, the results are even stronger, and a comparison of the number of unassociated $\gamma$-ray sources with the population of MSPs indicates that no more than 25\% of MSPs can be radio-quiet~\citep{TheFermi-LAT:2013ssa}. This implies that the lack of detected MSPs in the ATNF catalog may itself put strong constraints on the number of $\gamma$-ray MSPs in the GC region. 

\begin{figure}
\includegraphics[width=250pt]{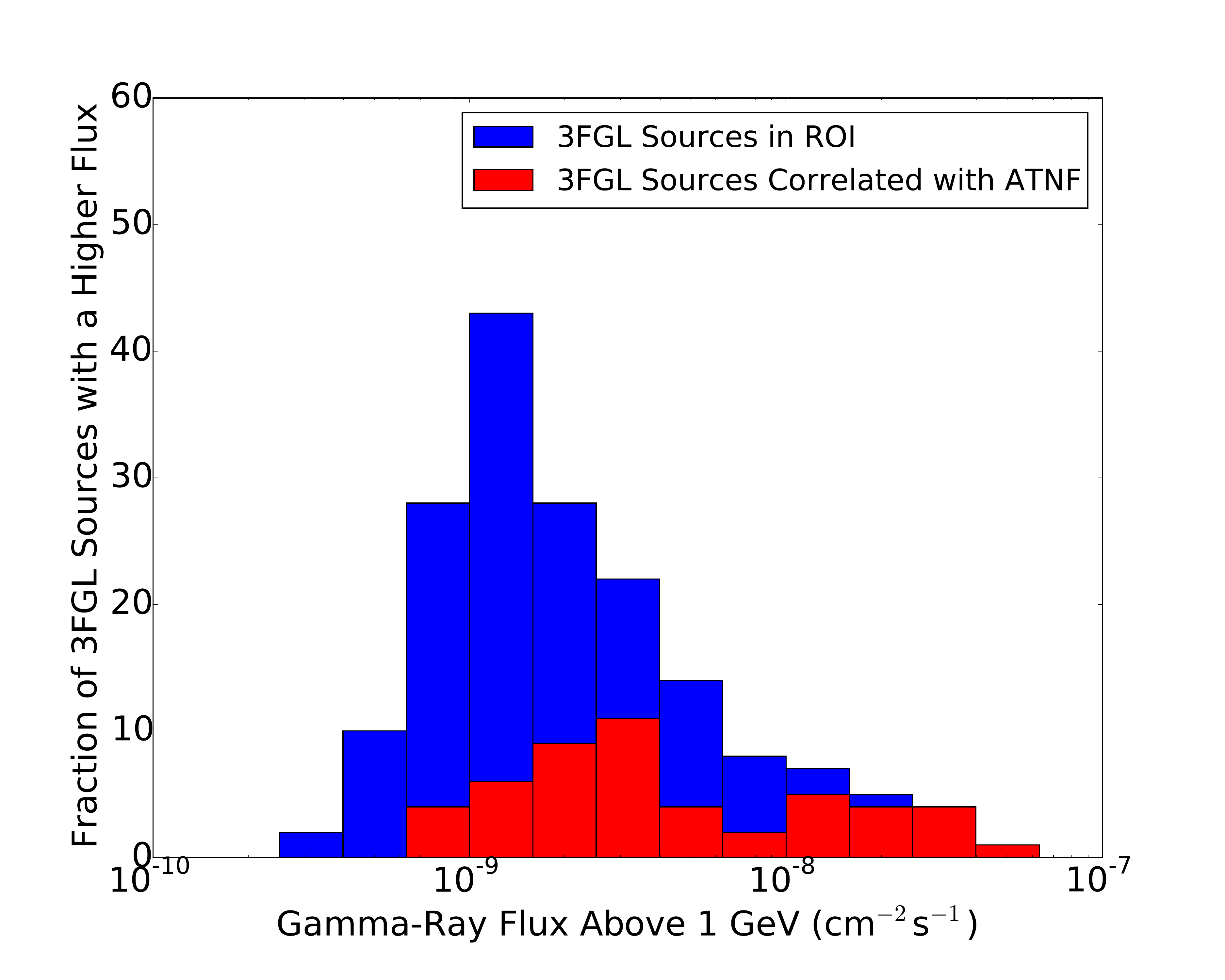}
\caption{\label{fig:ATNF_Correspondence} The $\gamma$-ray flux distribution of 3FGL sources in the GC and sidebands region, compared to the sub-sample of systems that is correlated with an ATNF pulsar. We observe a trend where pulsars with high $\gamma$-ray fluxes are more likely to be correlated with an ATNF pulsar. Several physical and systematic mechanisms are discussed in the text.}
\end{figure}

However, in Figure~\ref{fig:ATNF_Correspondence} we analyze the probability of associations as a function of $\gamma$-ray flux for the ATNF pulsars in the ROI under consideration, and do observe a trend. Specifically, we find that only 10\%$^{+7.9\%}_{-4.8\%}$ of 3FGL sources with a flux smaller than 10$^{-9}$cm$^{-2}$~s$^{-1}$ are correlated with ATNF pulsars, while 35\%$^{+5.9\%}_{-5.1\%}$ of 3FGL sources with higher luminosities show such a correlation. This may be due to physical effects, such as the distance-based correlation between $\gamma$-ray and radio fluxes, or a correlation between $\gamma$-ray fluxes and the fraction of pulsars which are radio quiet. We note that the second effect is thought to be due to the orientation of the pulsar spin axis with respect to the Earth, and is likely to be averaged out over any large pulsar population.

The correlation may also fail if a new population of $\gamma$-ray emitters (such as high-mass binaries) dominates the population of low-flux $\gamma$-ray emitters. The substantial fraction of high-TS hotspots at mirrored pulsar locations in this analysis also hints that a substantial fraction of low-flux point sources in the GC region may, in fact, be artifacts due to diffuse mismodeling. Finally, the correlation may be due to observational effects, such as errors in calculating the position of low-flux $\gamma$-ray point sources, and may be relieved if upcoming radio follow-ups observe radio pulsars at the positions of more recently detected low-flux $\gamma$-ray emitters.

Even if the correlation between $\gamma$-ray pulsars and ATNF sources falls to $\sim$10\% at lower $\gamma$-ray luminosities, our analysis would still disfavor $\gamma$-ray pulsars as a contributor to the GC excess. However, at present it is difficult to rule out a $<$5\% correlation between sub-threshold $\gamma$-ray pulsars and ATNF sources, which would make any application of our ATNF results to the total pulsar population in the GC region difficult to ascertain.

\section{Discussion and Conclusions}
\label{sec:conclusions}

Our analysis indicates that any pulsar population which contributes to the $\gamma$-ray excess is not represented in the ATNF survey of radio pulsars. There are two potential conclusions to draw from this study. Conservatively, we conclude that ATNF pulsars that are not already associated with 3FGL point sources do not contribute any significant $\gamma$-ray emission in the GC region. This result is unfortunate, as a multiwavelength correlation between $\gamma$-ray hotspots in the GC and the ATNF catalog would serve as smoking gun evidence for the pulsar origin of the GC excess.

Our analysis also indicates that pulsars drawn from similar populations as those appearing in the ATNF catalog are unlikely to explain the GC excess. The 41 ATNF pulsars with distance measurements consistent with a GC population appear to produce no sizeable $\gamma$-ray emission ($<$3.65 sources with TS$>$20 correspond to true $\gamma$-ray pulsars). Thus, any population consisting of hundreds of $\gamma$-ray bright pulsars must have an extremely small overlap with ATNF sources. This conclusion strongly disfavors the scenario put forth by \citep{O'Leary:2015gfa}, where a young pulsar population in the GC produces the GC excess. However, the small population of MSPs in the ATNF survey makes this conclusion more easy to avoid in scenarios where the GC excess is attributed to a population of MSPs. Systematic follow-up studies of MSP radio sensitivies throughout the GC region are necessary in order to determine whether a substantial MSP population may hide below the detectability threshold of current instrumentation.

Finally, we note that this work does not elucidate the resilience of the GC excess to the addition of point sources near the GC region. While it is likely that numerous $\gamma$-ray point sources contribute to the GC environment, it is currently unclear whether these missing sources are currently attributed to the GC excess, or soaked up by galactic diffuse models. In an upcoming study, we plan to revisit analyses of the GC excess in models with substantial populations of currently undetected point sources, in order to determine whether any missing point source population is indeed capable of explaining the GC excess~\citep{nextpaper}.

\acknowledgments
TL thanks Dan Hooper for many useful comments throughout the production of this manuscript as well as Eric Carlson, Annika Peter and Christoph Weniger for helpful comments. TL is supported by the National Aeronautics and Space Administration through Einstein Postdoctoral Fellowship Award Number PF3-140110. This work made use of computing resources and support provided by the Research Computing Center at the University of Chicago.

\bibliography{pulsars} 

\begin{thebibliography}{37}
\expandafter\ifx\csname natexlab\endcsname\relax\def\natexlab#1{#1}\fi
\expandafter\ifx\csname bibnamefont\endcsname\relax
  \def\bibnamefont#1{#1}\fi
\expandafter\ifx\csname bibfnamefont\endcsname\relax
  \def\bibfnamefont#1{#1}\fi
\expandafter\ifx\csname citenamefont\endcsname\relax
  \def\citenamefont#1{#1}\fi
\expandafter\ifx\csname url\endcsname\relax
  \def\url#1{\texttt{#1}}\fi
\expandafter\ifx\csname urlprefix\endcsname\relax\def\urlprefix{URL }\fi
\providecommand{\bibinfo}[2]{#2}
\providecommand{\eprint}[2][]{\url{#2}}

\bibitem[{\citenamefont{Goodenough and Hooper}(2009)}]{Goodenough:2009gk}
\bibinfo{author}{\bibfnamefont{L.}~\bibnamefont{Goodenough}} \bibnamefont{and}
  \bibinfo{author}{\bibfnamefont{D.}~\bibnamefont{Hooper}}
  (\bibinfo{year}{2009}), \eprint{0910.2998}.

\bibitem[{\citenamefont{Hooper and Goodenough}(2011)}]{Hooper:2010mq}
\bibinfo{author}{\bibfnamefont{D.}~\bibnamefont{Hooper}} \bibnamefont{and}
  \bibinfo{author}{\bibfnamefont{L.}~\bibnamefont{Goodenough}},
  \bibinfo{journal}{Phys.Lett.} \textbf{\bibinfo{volume}{B697}},
  \bibinfo{pages}{412} (\bibinfo{year}{2011}), \eprint{1010.2752}.

\bibitem[{\citenamefont{Hooper and Linden}(2011)}]{Hooper:2011ti}
\bibinfo{author}{\bibfnamefont{D.}~\bibnamefont{Hooper}} \bibnamefont{and}
  \bibinfo{author}{\bibfnamefont{T.}~\bibnamefont{Linden}},
  \bibinfo{journal}{Phys.Rev.} \textbf{\bibinfo{volume}{D84}},
  \bibinfo{pages}{123005} (\bibinfo{year}{2011}), \eprint{1110.0006}.

\bibitem[{\citenamefont{Abazajian and Kaplinghat}(2012)}]{Abazajian:2012pn}
\bibinfo{author}{\bibfnamefont{K.~N.} \bibnamefont{Abazajian}}
  \bibnamefont{and}
  \bibinfo{author}{\bibfnamefont{M.}~\bibnamefont{Kaplinghat}},
  \bibinfo{journal}{Phys.Rev.} \textbf{\bibinfo{volume}{D86}},
  \bibinfo{pages}{083511} (\bibinfo{year}{2012}), \eprint{1207.6047}.

\bibitem[{\citenamefont{Gordon and Macias}(2013)}]{Gordon:2013vta}
\bibinfo{author}{\bibfnamefont{C.}~\bibnamefont{Gordon}} \bibnamefont{and}
  \bibinfo{author}{\bibfnamefont{O.}~\bibnamefont{Macias}},
  \bibinfo{journal}{Phys.Rev.} \textbf{\bibinfo{volume}{D88}},
  \bibinfo{pages}{083521} (\bibinfo{year}{2013}), \eprint{1306.5725}.

\bibitem[{\citenamefont{Macias and Gordon}(2014)}]{Macias:2013vya}
\bibinfo{author}{\bibfnamefont{O.}~\bibnamefont{Macias}} \bibnamefont{and}
  \bibinfo{author}{\bibfnamefont{C.}~\bibnamefont{Gordon}},
  \bibinfo{journal}{Phys.Rev.} \textbf{\bibinfo{volume}{D89}},
  \bibinfo{pages}{063515} (\bibinfo{year}{2014}), \eprint{1312.6671}.

\bibitem[{\citenamefont{Abazajian et~al.}(2014)\citenamefont{Abazajian, Canac,
  Horiuchi, and Kaplinghat}}]{Abazajian:2014fta}
\bibinfo{author}{\bibfnamefont{K.~N.} \bibnamefont{Abazajian}},
  \bibinfo{author}{\bibfnamefont{N.}~\bibnamefont{Canac}},
  \bibinfo{author}{\bibfnamefont{S.}~\bibnamefont{Horiuchi}}, \bibnamefont{and}
  \bibinfo{author}{\bibfnamefont{M.}~\bibnamefont{Kaplinghat}},
  \bibinfo{journal}{Phys.Rev.} \textbf{\bibinfo{volume}{D90}},
  \bibinfo{pages}{023526} (\bibinfo{year}{2014}), \eprint{1402.4090}.

\bibitem[{\citenamefont{Hooper and Slatyer}(2013)}]{Hooper:2013rwa}
\bibinfo{author}{\bibfnamefont{D.}~\bibnamefont{Hooper}} \bibnamefont{and}
  \bibinfo{author}{\bibfnamefont{T.~R.} \bibnamefont{Slatyer}},
  \bibinfo{journal}{Phys.Dark Univ.} \textbf{\bibinfo{volume}{2}},
  \bibinfo{pages}{118} (\bibinfo{year}{2013}), \eprint{1302.6589}.

\bibitem[{\citenamefont{Daylan et~al.}(2014)\citenamefont{Daylan, Finkbeiner,
  Hooper, Linden, Portillo et~al.}}]{Daylan:2014rsa}
\bibinfo{author}{\bibfnamefont{T.}~\bibnamefont{Daylan}},
  \bibinfo{author}{\bibfnamefont{D.~P.} \bibnamefont{Finkbeiner}},
  \bibinfo{author}{\bibfnamefont{D.}~\bibnamefont{Hooper}},
  \bibinfo{author}{\bibfnamefont{T.}~\bibnamefont{Linden}},
  \bibinfo{author}{\bibfnamefont{S.~K.~N.} \bibnamefont{Portillo}},
  \bibnamefont{et~al.} (\bibinfo{year}{2014}), \eprint{1402.6703}.

\bibitem[{\citenamefont{Calore et~al.}(2015)\citenamefont{Calore, Cholis, and
  Weniger}}]{Calore:2014xka}
\bibinfo{author}{\bibfnamefont{F.}~\bibnamefont{Calore}},
  \bibinfo{author}{\bibfnamefont{I.}~\bibnamefont{Cholis}}, \bibnamefont{and}
  \bibinfo{author}{\bibfnamefont{C.}~\bibnamefont{Weniger}},
  \bibinfo{journal}{JCAP} \textbf{\bibinfo{volume}{1503}}, \bibinfo{pages}{038}
  (\bibinfo{year}{2015}), \eprint{1409.0042}.

\bibitem[{\citenamefont{Berlin et~al.}(2014)\citenamefont{Berlin, Hooper, and
  McDermott}}]{Berlin:2014tja}
\bibinfo{author}{\bibfnamefont{A.}~\bibnamefont{Berlin}},
  \bibinfo{author}{\bibfnamefont{D.}~\bibnamefont{Hooper}}, \bibnamefont{and}
  \bibinfo{author}{\bibfnamefont{S.~D.} \bibnamefont{McDermott}},
  \bibinfo{journal}{Phys.Rev.} \textbf{\bibinfo{volume}{D89}},
  \bibinfo{pages}{115022} (\bibinfo{year}{2014}), \eprint{1404.0022}.

\bibitem[{\citenamefont{Agrawal et~al.}(2014)\citenamefont{Agrawal, Batell,
  Hooper, and Lin}}]{Agrawal:2014una}
\bibinfo{author}{\bibfnamefont{P.}~\bibnamefont{Agrawal}},
  \bibinfo{author}{\bibfnamefont{B.}~\bibnamefont{Batell}},
  \bibinfo{author}{\bibfnamefont{D.}~\bibnamefont{Hooper}}, \bibnamefont{and}
  \bibinfo{author}{\bibfnamefont{T.}~\bibnamefont{Lin}},
  \bibinfo{journal}{Phys.Rev.} \textbf{\bibinfo{volume}{D90}},
  \bibinfo{pages}{063512} (\bibinfo{year}{2014}), \eprint{1404.1373}.

\bibitem[{\citenamefont{Alves et~al.}(2014)\citenamefont{Alves, Profumo,
  Queiroz, and Shepherd}}]{Alves:2014yha}
\bibinfo{author}{\bibfnamefont{A.}~\bibnamefont{Alves}},
  \bibinfo{author}{\bibfnamefont{S.}~\bibnamefont{Profumo}},
  \bibinfo{author}{\bibfnamefont{F.~S.} \bibnamefont{Queiroz}},
  \bibnamefont{and} \bibinfo{author}{\bibfnamefont{W.}~\bibnamefont{Shepherd}},
  \bibinfo{journal}{Phys.Rev.} \textbf{\bibinfo{volume}{D90}},
  \bibinfo{pages}{115003} (\bibinfo{year}{2014}), \eprint{1403.5027}.

\bibitem[{\citenamefont{Abdullah et~al.}(2014)\citenamefont{Abdullah, DiFranzo,
  Rajaraman, Tait, Tanedo et~al.}}]{Abdullah:2014lla}
\bibinfo{author}{\bibfnamefont{M.}~\bibnamefont{Abdullah}},
  \bibinfo{author}{\bibfnamefont{A.}~\bibnamefont{DiFranzo}},
  \bibinfo{author}{\bibfnamefont{A.}~\bibnamefont{Rajaraman}},
  \bibinfo{author}{\bibfnamefont{T.~M.} \bibnamefont{Tait}},
  \bibinfo{author}{\bibfnamefont{P.}~\bibnamefont{Tanedo}},
  \bibnamefont{et~al.}, \bibinfo{journal}{Phys.Rev.}
  \textbf{\bibinfo{volume}{D90}}, \bibinfo{pages}{035004}
  (\bibinfo{year}{2014}), \eprint{1404.6528}.

\bibitem[{\citenamefont{Izaguirre et~al.}(2014)\citenamefont{Izaguirre,
  Krnjaic, and Shuve}}]{Izaguirre:2014vva}
\bibinfo{author}{\bibfnamefont{E.}~\bibnamefont{Izaguirre}},
  \bibinfo{author}{\bibfnamefont{G.}~\bibnamefont{Krnjaic}}, \bibnamefont{and}
  \bibinfo{author}{\bibfnamefont{B.}~\bibnamefont{Shuve}},
  \bibinfo{journal}{Phys.Rev.} \textbf{\bibinfo{volume}{D90}},
  \bibinfo{pages}{055002} (\bibinfo{year}{2014}), \eprint{1404.2018}.

\bibitem[{\citenamefont{Carlson and Profumo}(2014)}]{Carlson:2014cwa}
\bibinfo{author}{\bibfnamefont{E.}~\bibnamefont{Carlson}} \bibnamefont{and}
  \bibinfo{author}{\bibfnamefont{S.}~\bibnamefont{Profumo}},
  \bibinfo{journal}{Phys.Rev.} \textbf{\bibinfo{volume}{D90}},
  \bibinfo{pages}{023015} (\bibinfo{year}{2014}), \eprint{1405.7685}.

\bibitem[{\citenamefont{Petrovic et~al.}(2014)\citenamefont{Petrovic, Serpico,
  and Zaharijas}}]{Petrovic:2014uda}
\bibinfo{author}{\bibfnamefont{J.}~\bibnamefont{Petrovic}},
  \bibinfo{author}{\bibfnamefont{P.~D.} \bibnamefont{Serpico}},
  \bibnamefont{and}
  \bibinfo{author}{\bibfnamefont{G.}~\bibnamefont{Zaharijas}},
  \bibinfo{journal}{JCAP} \textbf{\bibinfo{volume}{1410}}, \bibinfo{pages}{052}
  (\bibinfo{year}{2014}), \eprint{1405.7928}.

\bibitem[{\citenamefont{Cholis et~al.}(2015)\citenamefont{Cholis, Evoli,
  Calore, Linden, Weniger, and Hooper}}]{Cholis:2015dea}
\bibinfo{author}{\bibfnamefont{I.}~\bibnamefont{Cholis}},
  \bibinfo{author}{\bibfnamefont{C.}~\bibnamefont{Evoli}},
  \bibinfo{author}{\bibfnamefont{F.}~\bibnamefont{Calore}},
  \bibinfo{author}{\bibfnamefont{T.}~\bibnamefont{Linden}},
  \bibinfo{author}{\bibfnamefont{C.}~\bibnamefont{Weniger}}, \bibnamefont{and}
  \bibinfo{author}{\bibfnamefont{D.}~\bibnamefont{Hooper}}
  (\bibinfo{year}{2015}), \eprint{1506.05119}.

\bibitem[{\citenamefont{Abazajian}(2011)}]{Abazajian:2010zy}
\bibinfo{author}{\bibfnamefont{K.~N.} \bibnamefont{Abazajian}},
  \bibinfo{journal}{JCAP} \textbf{\bibinfo{volume}{1103}}, \bibinfo{pages}{010}
  (\bibinfo{year}{2011}), \eprint{1011.4275}.

\bibitem[{\citenamefont{Yuan and Ioka}(2015)}]{Yuan:2014yda}
\bibinfo{author}{\bibfnamefont{Q.}~\bibnamefont{Yuan}} \bibnamefont{and}
  \bibinfo{author}{\bibfnamefont{K.}~\bibnamefont{Ioka}},
  \bibinfo{journal}{Astrophys.J.} \textbf{\bibinfo{volume}{802}},
  \bibinfo{pages}{124} (\bibinfo{year}{2015}), \eprint{1411.4363}.

\bibitem[{\citenamefont{Petrović et~al.}(2015)\citenamefont{Petrović,
  Serpico, and Zaharijas}}]{Petrovic:2014xra}
\bibinfo{author}{\bibfnamefont{J.}~\bibnamefont{Petrović}},
  \bibinfo{author}{\bibfnamefont{P.~D.} \bibnamefont{Serpico}},
  \bibnamefont{and}
  \bibinfo{author}{\bibfnamefont{G.}~\bibnamefont{Zaharijas}},
  \bibinfo{journal}{JCAP} \textbf{\bibinfo{volume}{1502}}, \bibinfo{pages}{023}
  (\bibinfo{year}{2015}), \eprint{1411.2980}.

\bibitem[{\citenamefont{O'Leary et~al.}(2015)\citenamefont{O'Leary, Kistler,
  Kerr, and Dexter}}]{O'Leary:2015gfa}
\bibinfo{author}{\bibfnamefont{R.~M.} \bibnamefont{O'Leary}},
  \bibinfo{author}{\bibfnamefont{M.~D.} \bibnamefont{Kistler}},
  \bibinfo{author}{\bibfnamefont{M.}~\bibnamefont{Kerr}}, \bibnamefont{and}
  \bibinfo{author}{\bibfnamefont{J.}~\bibnamefont{Dexter}}
  (\bibinfo{year}{2015}), \eprint{1504.02477}.

\bibitem[{\citenamefont{Hooper et~al.}(2013)\citenamefont{Hooper, Cholis,
  Linden, Siegal-Gaskins, and Slatyer}}]{Hooper:2013nhl}
\bibinfo{author}{\bibfnamefont{D.}~\bibnamefont{Hooper}},
  \bibinfo{author}{\bibfnamefont{I.}~\bibnamefont{Cholis}},
  \bibinfo{author}{\bibfnamefont{T.}~\bibnamefont{Linden}},
  \bibinfo{author}{\bibfnamefont{J.}~\bibnamefont{Siegal-Gaskins}},
  \bibnamefont{and} \bibinfo{author}{\bibfnamefont{T.}~\bibnamefont{Slatyer}},
  \bibinfo{journal}{Phys.Rev.} \textbf{\bibinfo{volume}{D88}},
  \bibinfo{pages}{083009} (\bibinfo{year}{2013}), \eprint{1305.0830}.

\bibitem[{\citenamefont{Cholis et~al.}(2014{\natexlab{a}})\citenamefont{Cholis,
  Hooper, and Linden}}]{Cholis:2014noa}
\bibinfo{author}{\bibfnamefont{I.}~\bibnamefont{Cholis}},
  \bibinfo{author}{\bibfnamefont{D.}~\bibnamefont{Hooper}}, \bibnamefont{and}
  \bibinfo{author}{\bibfnamefont{T.}~\bibnamefont{Linden}}
  (\bibinfo{year}{2014}{\natexlab{a}}), \eprint{1407.5583}.

\bibitem[{\citenamefont{Cholis et~al.}(2014{\natexlab{b}})\citenamefont{Cholis,
  Hooper, and Linden}}]{Cholis:2014lta}
\bibinfo{author}{\bibfnamefont{I.}~\bibnamefont{Cholis}},
  \bibinfo{author}{\bibfnamefont{D.}~\bibnamefont{Hooper}}, \bibnamefont{and}
  \bibinfo{author}{\bibfnamefont{T.}~\bibnamefont{Linden}}
  (\bibinfo{year}{2014}{\natexlab{b}}), \eprint{1407.5625}.

\bibitem[{\citenamefont{Brandt and Kocsis}(2015)}]{Brandt:2015ula}
\bibinfo{author}{\bibfnamefont{T.~D.} \bibnamefont{Brandt}} \bibnamefont{and}
  \bibinfo{author}{\bibfnamefont{B.}~\bibnamefont{Kocsis}}
  (\bibinfo{year}{2015}), \eprint{1507.05616}.

\bibitem[{\citenamefont{Bartels et~al.}(2015)\citenamefont{Bartels,
  Krishnamurthy, and Weniger}}]{Bartels:2015aea}
\bibinfo{author}{\bibfnamefont{R.}~\bibnamefont{Bartels}},
  \bibinfo{author}{\bibfnamefont{S.}~\bibnamefont{Krishnamurthy}},
  \bibnamefont{and} \bibinfo{author}{\bibfnamefont{C.}~\bibnamefont{Weniger}}
  (\bibinfo{year}{2015}), \eprint{1506.05104}.

\bibitem[{The(2015)}]{TheFermi-LAT:2015hja}
 (\bibinfo{year}{2015}), \eprint{1501.02003}.

\bibitem[{\citenamefont{Lee et~al.}(2015)\citenamefont{Lee, Lisanti, Safdi,
  Slatyer, and Xue}}]{Lee:2015fea}
\bibinfo{author}{\bibfnamefont{S.~K.} \bibnamefont{Lee}},
  \bibinfo{author}{\bibfnamefont{M.}~\bibnamefont{Lisanti}},
  \bibinfo{author}{\bibfnamefont{B.~R.} \bibnamefont{Safdi}},
  \bibinfo{author}{\bibfnamefont{T.~R.} \bibnamefont{Slatyer}},
  \bibnamefont{and} \bibinfo{author}{\bibfnamefont{W.}~\bibnamefont{Xue}}
  (\bibinfo{year}{2015}), \eprint{1506.05124}.

\bibitem[{\citenamefont{Manchester et~al.}(2005)\citenamefont{Manchester,
  Hobbs, Teoh, and Hobbs}}]{Manchester:2004bp}
\bibinfo{author}{\bibfnamefont{R.~N.} \bibnamefont{Manchester}},
  \bibinfo{author}{\bibfnamefont{G.~B.} \bibnamefont{Hobbs}},
  \bibinfo{author}{\bibfnamefont{A.}~\bibnamefont{Teoh}}, \bibnamefont{and}
  \bibinfo{author}{\bibfnamefont{M.}~\bibnamefont{Hobbs}},
  \bibinfo{journal}{Astron. J.} \textbf{\bibinfo{volume}{129}},
  \bibinfo{pages}{1993} (\bibinfo{year}{2005}), \eprint{astro-ph/0412641}.

\bibitem[{\citenamefont{Taylor and Cordes}(1993)}]{Taylor:1993my}
\bibinfo{author}{\bibfnamefont{J.~H.} \bibnamefont{Taylor}} \bibnamefont{and}
  \bibinfo{author}{\bibfnamefont{J.~M.} \bibnamefont{Cordes}},
  \bibinfo{journal}{Astrophys. J.} \textbf{\bibinfo{volume}{411}},
  \bibinfo{pages}{674} (\bibinfo{year}{1993}).

\bibitem[{\citenamefont{{Cordes} and {Lazio}}(2002)}]{2002astro.ph..7156C}
\bibinfo{author}{\bibfnamefont{J.~M.} \bibnamefont{{Cordes}}} \bibnamefont{and}
  \bibinfo{author}{\bibfnamefont{T.~J.~W.} \bibnamefont{{Lazio}}},
  \bibinfo{journal}{ArXiv Astrophysics e-prints}  (\bibinfo{year}{2002}),
  \eprint{astro-ph/0207156}.

\bibitem[{\citenamefont{Ackermann et~al.}(2014)}]{Ackermann:2013yva}
\bibinfo{author}{\bibfnamefont{M.}~\bibnamefont{Ackermann}}
  \bibnamefont{et~al.} (\bibinfo{collaboration}{Fermi-LAT}),
  \bibinfo{journal}{Phys.Rev.} \textbf{\bibinfo{volume}{D89}},
  \bibinfo{pages}{042001} (\bibinfo{year}{2014}), \eprint{1310.0828}.

\bibitem[{\citenamefont{Hooper and Linden}(2015)}]{Hooper:2015ula}
\bibinfo{author}{\bibfnamefont{D.}~\bibnamefont{Hooper}} \bibnamefont{and}
  \bibinfo{author}{\bibfnamefont{T.}~\bibnamefont{Linden}}
  (\bibinfo{year}{2015}), \eprint{1503.06209}.

\bibitem[{\citenamefont{{Linden}}(2015)}]{nextpaper}
\bibinfo{author}{\bibfnamefont{T.}~\bibnamefont{{Linden}}}
  (\bibinfo{year}{2015}).

\bibitem[{\citenamefont{Marelli et~al.}(2015)\citenamefont{Marelli, Mignani,
  De~Luca, Saz~Parkinson, Salvetti, Den~Hartog, and Wolff}}]{Marelli:2015vsa}
\bibinfo{author}{\bibfnamefont{M.}~\bibnamefont{Marelli}},
  \bibinfo{author}{\bibfnamefont{R.~P.} \bibnamefont{Mignani}},
  \bibinfo{author}{\bibfnamefont{A.}~\bibnamefont{De~Luca}},
  \bibinfo{author}{\bibfnamefont{P.~M.} \bibnamefont{Saz~Parkinson}},
  \bibinfo{author}{\bibfnamefont{D.}~\bibnamefont{Salvetti}},
  \bibinfo{author}{\bibfnamefont{P.~R.} \bibnamefont{Den~Hartog}},
  \bibnamefont{and} \bibinfo{author}{\bibfnamefont{M.~T.} \bibnamefont{Wolff}},
  \bibinfo{journal}{Astrophys. J.} \textbf{\bibinfo{volume}{802}},
  \bibinfo{pages}{78} (\bibinfo{year}{2015}), \eprint{1501.06215}.

\bibitem[{\citenamefont{Abdo et~al.}(2013)}]{TheFermi-LAT:2013ssa}
\bibinfo{author}{\bibfnamefont{A.~A.} \bibnamefont{Abdo}} \bibnamefont{et~al.}
  (\bibinfo{collaboration}{Fermi-LAT}), \bibinfo{journal}{Astrophys. J. Suppl.}
  \textbf{\bibinfo{volume}{208}}, \bibinfo{pages}{17} (\bibinfo{year}{2013}),
  \eprint{1305.4385}.

\end{thebibliography}

\begin{appendix}

\begin{figure*}
\includegraphics[width=500pt]{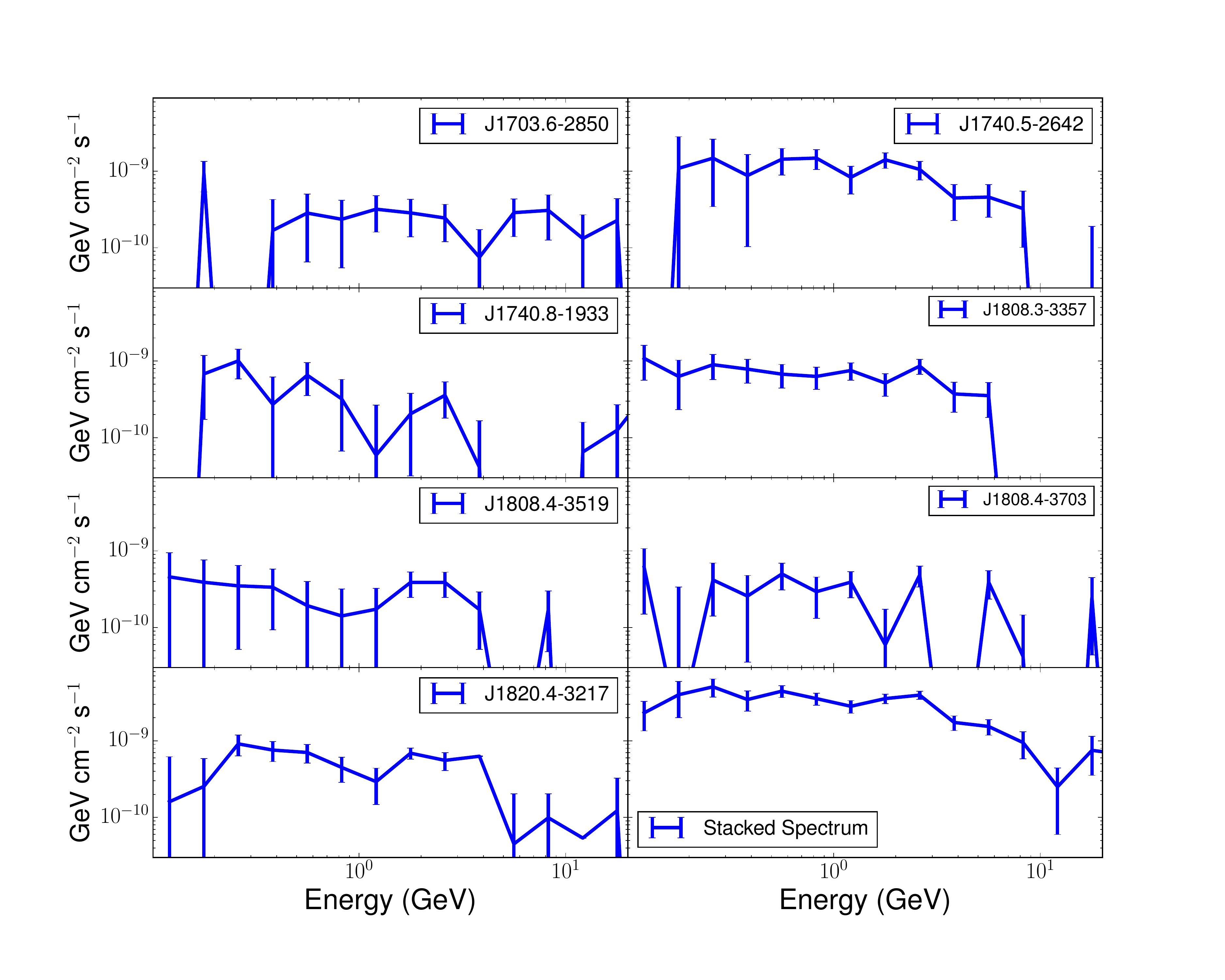}
\caption{\label{fig:bartels_3FGL} The $\gamma$-ray spectrum obtained for 7 of the $\gamma$-ray point sources indicated as possible pulsars in the analysis of \citep{Bartels:2015aea}. The proceedure utilized in this paper provides a more detailed spectral model of $\gamma$-ray sources compared to that utilized for the 3FGL catalog. We find that some sub-population of these sources have spectral characteristics consistent with MSPs, while others clearly have excessive levels of low-energy emission.} 
\end{figure*}

\section{An Analysis of the Gamma-Ray Point Sources Identified as Possible MSPs by Bartels et al. (2015)}

An analysis by \citep{Bartels:2015aea} analyzed the point sources in the GC region, and indicated 13 known 3FGL sources that have  MSP-like spectra, 7 of which lie within 10$^\circ$ of the GC (and are covered within the ROI of our analysis). While the 3FGL spectral measurements are consistent with a MSP origin for these systems, we note that this is based on intensity measurements of each source in only 5 energy bins, utilizing 4 years of $\gamma$-ray data~\citep{TheFermi-LAT:2015hja}. We examine this finding by utilizing the spectral fitting algorithm described in Section~\ref{sec:fermimodels} in order to calculate the spectrum of each source in much greater detail. In Figure~\ref{fig:bartels_3FGL} we show the resulting spectra obtained for each 3FGL source, finding varied results. Some 3FGL sources, (notably J1808.4-3519 and J1703.6-2850) appear to have spectra which are consistent with the MSP population, while othes (such as J1808.3-3357 and J1820.4-3217) appear to have power-law best fit spectra and contain significantly more low-energy emission compared to MSP models. In the bottom right of Figure~\ref{fig:bartels_3FGL} we show the stacked spectrum from the seven $\gamma$-ray sources, showing emission which is generally consistent with power-law behavior. It remains possible that a subset of these systems does in fact have an MSP origin, a result which would not be in tension either MSP or dark matter explanations for the $\gamma$-ray excess. We note that none of these 13 sources lie within 0.1$^\circ$ of any ATNF pulsar, but two sources (of the 6 outside our 10$^\circ$ ROI), lie at $\sim$0.25$^\circ$ from a known ATNF pulsar (3FGL J1744.8-1557 and 3FGL J1837.3-2403).

In theory, a similar test could be run on the position of $\gamma$-ray hotspots identified by \citep{Lee:2015fea}. However, it is practically difficult due to the low angular resolution (0.5$^\circ$) of the non-Poissonian template fitting algorithm used in that analysis. The poor positioning of these $\gamma$-ray hotspots makes it difficult to accurately place point sources at their center, potentially introducing significant spectral distortions from the positioning errors. Thus, we do not attempt such an analysis here. 

\end{appendix}

\end{document}